# Effects Of Zoning On Housing Option Value

Prathamesh Muzumdar, Illinois State University, Normal, USA

**ABSTRACT**

*The research explores the subject of zoning effect on price value of a house in a certain designated zone. Zoning is defined as a land under planned use. This results in price value of the land. As the area of the land decreases an increase in price occurs. This research does not specify the magnitude of a certain unit by which a housing price will increase or decrease but it estimates the housing option value. This research helps the policy makers to improve long term policy decisions. This research will showcase zoning effects on housing option value for the specific zones in the city of Normal. The observations are taken for different parcels of total neighborhoods in the city of Normal. This research will use hedonic model for conducting a regression analysis on the subject. The hypothesis taken into consideration before starting the research is to prove that more 50 percent variation in the housing price is affected by the zoning characteristics. This research will try to prove the hypothesis by comparing the assessed values with the predicted values from the model.*

**INTRODUCTION**

Housing price depends upon multiple variables. These variables contain both zoning as well as physical variables and other factors. Zoning variables, as shown in Table 1, are defined as:

R1 A:    Single family residence district.
R1 B:    Single family residence district with multifamily dwellings.
R2:      Mixed residence district.
S2:      Public land and institutional districts.

Physical Variables are defined as:
- log(lotsqfeet)
- log(lotdimb)
- log(lotdima)
- lot(totbldgft)
- lot(bathrooms)
- Age,
- Age^2
- Condition
- Tax Rate

Zoning differentiates land as per its use. Different zoning areas have different conditions. Somebody planning to do alteration has to take permissions doing these alterations. Thus, zoning affects the value of prices. Zoning is done by Governments and in USA it is done by federal agency. Zoning restricts the use of land in random way. E.g. the land available for building single family houses was 150 lots per parcel, after enacting zoning the land available would be 110 lots per parcel. This results in rise of prices in a certain zones. The change in property value due to zoning is known as option value. Zoning thus helps to use a certain value of land for its designated use. This research will use hedonic model to analyze how zoning will affect the housing value. The research will develop two models for developed and underdeveloped areas. The logarithmic value of house will be calculated using multiple variables. The most important part of the research would be the usage of different zone and different neighborhood areas of the city of Normal.

The relationship between different types of zones and its effects on the price of the house has been studied often because of the increasing urban planning by many researches. In this research value has been considered as dependant variable. The zoning effect will result in change in option value/ price of the house. The research has been constricted to the zoning patterns of the city of Normal. The primary purpose of zoning is to segregate land as per the design usage. Different zones are created for urban planning. Local governments are responsible for zoning. In the city of Normal zoning is used in the criterion of R1 A, R1 B, R2 and S2 which are most probably zone as the city constitutes of residential area. This zoning is done by Normal township assessor. Also the price valuation of houses differs as per different location. The locations near the basic amenities are prone to higher prices as compared to the areas with fewer amenities.

**LITERATURE REVIEW**

Lawrence Lai Wai Chang (1994) defined zoning with respect to city of Liverpool, UK in his thesis "The economics of land use zoning". In his thesis, zoning was described both advantageous and disadvantageous. Disadvantages were found to be failure to increase the efficiency for a proportion of land to its use. In spite of the back drop, zoning has proved to its practicality in USA as land availability is greater as compared to that of UK. This proves that zoning is effective way of urban planning over a greater stretch of land. Glasear and Gyaurko estimated changes in housing price resulting in housing affordability which results due to zoning impact. They have argued in their research that housing prices differ in different area as per zoning pattern. The pattern change occurs for different region of different states. Due to this housing price in California is greater as compared to other states.

Zoning in USA is of 4 types: Euclidian, Performance, Incentive, Design Based.

1. Euclidean:

    It is used mostly in small towns and large cities. Segregation of land is done on specified geographic district and dimensional standards stipulating limitations on development activity within each type of district.

2. Performance:

    It is performance based or goal oriented criteria to establish review parameters for proposed development projects. This type of zoning is intended to provide flexibility, rationality, transparency and accountability avoiding arbitrariness of the Euclidian approach and better accommodating market principles and private property rights with environmental protection.

3. Incentive:

    Implemented first in New York and Chicago, it is intended to provide a reward based system to encourage development that meets established urban development goals. Incentive zoning allows developers more density in exchange for community improvements. An increase in density encourages high density development supportive of compact development. In exchange, the developer would be encouraged to include some community improvements in their projects. Community improvements may include additional open space, affordable housing, special building features, or public art.

4. Design based (Form based):

    A form-based code (FBC) is a means of regulating development to achieve a specific urban form. Form-based codes create a predictable public realm by controlling physical form primarily, with a lesser focus on

land use, through city or county regulations. Form-based codes are a new response to the modern challenges of urban sprawl, deterioration of historic neighborhoods, and neglect of pedestrian safety in new development. Tradition has declined as a guide to development patterns, and the widespread adoption by cities of single-use zoning regulations has discouraged compact, walkable urbanism. Form-based codes are a tool to address these deficiencies, and to provide local governments the regulatory means to achieve development objectives with greater certainty.

**HEDONIC MODEL**

In a hedonic regression, the economist attempts to consistently estimate the relationship between prices and product attributes in a differentiated product market. The regression coefficients are commonly referred to as implicit (or hedonic) prices, which can be interpreted as the effect on the market price of increasing a particular product attribute while holding the other attributes fixed. Given utility-maximizing behavior, the consumer's marginal willingness to pay for a small change in a particular attribute can be inferred directly from an estimate of its implicit price; moreover, these implicit prices can be used to recover marginal willingness to pay functions for use in valuing larger changes in attributes (Rosen, 1974).

Hedonic regressions suffer from a number of well-known problems. Foremost among them, the economist is unlikely to directly observe all product characteristics that are relevant to consumers, and these omitted variables may lead to biased estimates of the implicit prices of the observed attributes. For example, in a house-price hedonic regression, the economist may observe the house's square-footage, lot size, and even the average education level in the neighborhood. However, many product attributes such as curb appeal, the quality of the landscaping and the crime rate may be unobserved by the econometrician. If these omitted attributes are correlated with the observed attributes, ordinary least squares estimates of the implicit prices will be biased.

**THE WORKING MODEL**

To evaluate the effect of zoning on residential property value, one must consider the effects of many different variables. Location characteristics are relevant in analyzing housing prices. Proximity to employment, recreation, roads, shopping centers, and many other agglomerations should increase the price of residential property as long as they do not produce negative externalities, in which case they should lower the value of the property. Public policy constraints and subsidies that include all types of land-use regulation and taxes will affect the value of one's property by increasing or decreasing the incentive to own it. One must also consider the influence of public good provision and the presence of amenities. They create desirability differences between pieces of property, thus creating differences in market value. The model that will be developed is applied to the city of Normal. Analysis will be conducted on a cross-sectional basis. The research considers a single model to study zoning with respect to developed residential property. For the model, the dependent variable is the natural log of total property value. Definitions for all variables are mentioned in Table 1 below.

The residential property model is:

Log(Value) = f(R1 A, R1B, R2, S2, log(lotsqfeet), log(lotdimb), log(lotdima), lot(totbldgft), lot(bathrooms), Age, Age^2, Condition, Tax Rate)

All zoning variables (R1 A, R1B, R2, and S2) are expected to change the value of residential property relative to leaving it unzoned. An increase in either land area or building area should increase the value of residential property, but the effect will decrease as either variable grows larger. The same effect is expected for an increase in the number of bathrooms. Age is a special variable. As a building's age increases, the value of the property is expected to decrease to a certain point, after which age becomes a valued amenity to prospective buyers. Hence, a quadratic form of the age variable was used to simulate its effects on residential property value. An increase in tax rate should decrease the value of the property since a higher tax burden will be capitalized into a lower price of housing (i.e., potential buyers would be willing to pay less in the face of higher taxes, all else equal). An increase in distance should also decrease the value of residential property due to relatively greater demand for property nearer to the central city, all else equal.

**Table 1. Definitions Of Variables**

| | |
|---|---|
| **Dependent Variable** | |
| Log(U1tfcash) | The total dollar value of the property (building and land) as assessed by Normal township assessor. |
| **Independent Variable** | |
| **Zoning Variables:** | |
| R1A | Equals 1 if zoned R1a, equals 0 otherwise. Primarily a single-family property in older areas with lots which do not meet the R1 standards. |
| R1 B | Equals 1 if zoned R1 B, equals 0 otherwise. Primarily a single-family and two-family residence district. Multi-family dwellings are permitted as conditional uses, and requirements for minimum lot size, ground floor area of structures and maximum height of buildings are somewhat less stringent than the requirements of the R1A zone. |
| R2 | Equals 1 if zoned R2, equals 0 otherwise. Permits all types of residential use, Including those parts of the city which are most densely built-up and contain a number of two- and multi-family dwellings. The minimum lot size requirements for dwellings in this district are lower to permit greater population densities close to the business and industrial areas. |
| S2 | Equals 1 if zoned S2, equals 0 otherwise. Permits only for public land and institutional district. |
| **Physical Variables:** | |
| Age | The age of any building included in the property. |
| Age^2 | Square of the age. |
| Log(lotsqfeet) | Lot square footage measured in square feet. |
| Log(lotdimb) | Lot Depth measured in feet. |
| Log(lotdima) | Lot width/Frontage measured in feet. |
| Log(totbldgft) | Total Building square feet. |
| Log(bathrooms) | Total number of bathrooms. |
| Condition | Equals 1 if rated as good condition (40% or higher), equals 0 if rated bad. |
| Tax Rate | The tax levy rate for the property (as a percentage of value). |

## DATA

Official data used by the Normal County Assessor's Office is obtained. The data set includes the entire population of residential and business property in Normal. Only residential property (Class II) is used in the analysis and any observation with missing data was eliminated. For the first model, the sample includes only those residential properties with a building (developed). For the model, the sample includes only those residential properties with a building (developed).

The data for the zoning variables was obtained from Ms. Mercy Davison, Town Planner, Department of Zoning, Town of Normal office (www.normal.org). The data consisted of 16 different zones which included the category of residential zones, mixed business and residential zones, business zones and special purpose zones. Out of these total zones 4 were selected as per their appropriate relevance to the model. Model designed only being for residential property value only requires the zones relevant to the residential property. The total data collected for the zoning characteristics was 12507 out of 12475 have been used for the model. The zone for particular house pin (property identification number) has been indicated by the numeric one while as for other zoned house it is indicated 0. In all these 4 zones constitute major residential property for the city of Normal. The non residential zones have been neglected to avoid numerous random error evolving out the model. The zoning pattern thus used indicates the nature of zone and its elasticity with respect to the housing price. Certain zones act in a positive manner giving rise to increase in price for accumulated residential property. Thus zoning has been found to have tremendous impact on the value of the house.

The data for the physical variable was obtained from Mr. Robert Cranston, Township assessor, Normal township office (www.normaltownship.org). The data consisted of 126 different variables out of which 9 variables were chosen for their appropriate relevancy with the model. These 9 variables have direct relationship with dependent variable price of the house. These variables either have a positive or negative linear relationship with the dependent variable where upon their t value indicates their nature. A data of 12507 was collected out of which 12475 was used for the model. The physical variable of tax rate, age and condition are subjected to change with time or sale of the house. Thus these physical variables have a tremendous impact on the value of the house.

## RESULTS

Table 2 shows the regression results for the model on developed residential properties. All co-efficient are significant at 10% significance level with the exception of total building feet and number of bathrooms which are not significant at 10% significance level. The R square is 0.89.

Table 2. Parameter Estimates For Developed Residential Property

| Variable | Parameter Estimate | Standard Error | t- Value |
|---|---|---|---|
| R1 A | 0.5592927 | 0.042187 | 13.26* |
| R1 B | 0.4670651 | 0.03866 | 12.08* |
| R 2 | 0.3999119 | 0.041335 | 9.67* |
| S2 | -10.68838 | 0.211277 | -50.59* |
| Log(lotsqfeet) | 0.1535171 | 0.044835 | 3.42* |
| Log(lotdimb) | -0.148944 | 0.062722 | -2.37* |
| Log(lotdima) | 0.2205652 | 0.043684 | 5.05* |
| Log(totbldgft) | 0.0385547 | 0.035657 | 1.08 |
| Log(bathrooms) | -0.00315 | 0.032486 | -0.10 |
| Age | -0.002379 | 0.000794 | -3.00* |
| Age^2 | 1.25516 | 3.909098 | 3.21* |
| condition | 0.1402772 | 0.046006 | 3.05* |
| taxrate | 0.2144761 | 0.088847 | 2.41* |
| F- Value | 402.3826 | | |
| R- Square | 0.8952 | | |
| Adjusted R- Square | 0.8930 | | |

\* Significant at ten percent level.

The R Square being the coefficient of determinant indicates that the 89 percent variation in the log(U1tfcash) which is the dependent variable ( assessed house price) can be explained by the variation in the independent variables which are R1 A, R1 B, R2, S2, lotdima, lotdimb, lotsqfeet, Age, Age^2, totbldgft, bathrooms, condition and taxrate. The changes taking place in the value of this independent variable will have a direct impact with 89 percent variation taking place in the dependent variable. A slight variation by one percent in any one this variable will cause 89 percent variation in the dependent variable. The house price variation is therefore a direct result of the variations its constituent variable factor.

Since total building feet and number of bathroom change as per need their statistical insignificance is not surprising. R1 A zoning with the parameter estimate of 0.55 has a greater impact on housing value than other zoning types. Also being second highly used with a density of 4192 the zone of R1 A is can affect the housing in greater manner compared to other zones. A 55 percent increase in value occurs when a residential property is zoned R1 A. R1 B has the second highest effect on housing price with a parameter estimate of 0.46. 46 percent increase in total value occurs when developed property is zoned R1 B. R1 B being the highly used zone with a density of 5219 has a parameter estimate of 0.46. Being the most highly used and the most popular for single family residence district it has an impact of 46 percent increase on the housing price in the residential zone. R1 B is then followed by R2 with 0.39 parameter estimate which indicates 39 percent increase in housing price when a developed property is zoned R2. R2 being a mixed resident district is third highly used zone with a density of 628. It has a weak negative relationship with the housing value where if a house is zoned in R2 there will be a decrease in price. S2 is the least impact full zone being a public land and institution district it has a negative effect on housing value. Also lot depth, number of bathrooms and age has a negative parameter estimate indicating negative effect on the value. Increase in lot depth will result in 14 percent decrease in the value of the house and increase in number of bathrooms will result in 0.3 percent decrease in the value of the house. Age is inversely proportional to housing price as the house becomes older its value falls by 0.2 percent.

Multivariate correlation matrix 1 for the independent variables of zoning indicate that the zone of R1 A and the zone R1 B are highly correlated with -0.45 as the value. A certain change in current norms of either in one of these zones will result in changing the estimates of the effects. The high collinearity showcases that R1 A tends to explain most of variance in the dependent leaving little variances to be explained by R1 B. Multivariate correlation matrix 2 for independent non physical variable indicates high correlation between condition and age with -0.34 as the value. As age and condition both reduce value of a house with time, it is of no surprise that their correlation is very high. Multivariate correlation matrix 3 for the independent physical variable indicates high correlation between lot square footage and lot width with 0.87 as the value. An increase or decrease in any of the variable of this two can result in changing the estimates of the effects.

Table 3. Multivariate Correlation Matrix

**Multivariate Correlation Matrix 1**

|  | Log(u1tfcash) | R1 A | R1 B | R 2 | S 2 |
|---|---|---|---|---|---|
| **Log(u1tfcash)** | 1.0000 | 0.2002 | 0.1257 | -0.0394 | -0.4685 |
| **R1 A** | 0.2002 | 1.0000 | -0.4512 | -0.2225 | -0.0496 |
| **R1 B** | 0.1257 | -0.4512 | 1.0000 | -0.2466 | -0.0549 |
| **R 2** | -0.0394 | -0.2225 | -0.2466 | 1.0000 | -0.0271 |
| **S 2** | -0.4685 | -0.0496 | -0.0549 | -0.0271 | 1.0000 |

**Multivariate Correlation Matrix 2**

|  | Log(u1tfcash) | taxrate | condition 2 | Age | Age^2 |
|---|---|---|---|---|---|
| **Log(u1tfcash)** | 1.0000 | 0.0387 | 0.1825 | -0.0714 | -0.0694 |
| **taxrate** | 0.0387 | 1.0000 | 0.0964 | -0.2808 | -0.2700 |
| **condition 2** | 0.1825 | 0.0964 | 1.0000 | -0.3422 | -0.3257 |
| **Age** | -0.0714 | -0.2808 | -0.3422 | 1.0000 | 0.9986 |
| **Age^2** | -0.0694 | -0.2700 | -0.3257 | 0.9986 | 1.0000 |

**Multivariate Correlation Matrix 3**

|  | Log(u1tfcash) | Log(lotdima) | Log(lotdimb) | Log(lotsqfeet) | Log(totbldgft) |
|---|---|---|---|---|---|
| **Log(u1tfcash)** | 1.0000 | 0.2950 | 0.0093 | 0.3477 | 0.0480 |
| **Log(lotdima)** | 0.2950 | 1.0000 | 0.3958 | 0.8706 | 0.0210 |
| **Log(lotdimb)** | 0.0093 | 0.3958 | 1.0000 | 0.7113 | 0.0631 |
| **Log(lotsqfeet)** | 0.3477 | 0.8706 | 0.7113 | 1.0000 | 0.0380 |
| **Log(totbldgft)** | 0.0480 | 0.0210 | 0.0631 | 0.0380 | 1.0000 |

Table 4. Descriptive Statistics For All Variables

**Developed Residential Property**

| Variable | Mean | Highest Value | Lowest Value | Value was yes |
|---|---|---|---|---|
| lotdima | 68.05 | 2384 | 20 | NA |
| lotdimb | 120.48 | 2032.5 | 77 | NA |
| lotsqfeet | 6063 | 250000 | 666 | NA |
| Age | - | - | - | NA |
| Age^2 | - | - | - | NA |
| Totbldgft | 2196.45 | 256609 | 645 | NA |
| Bathrooms | 3.57 | 336 | 1 | NA |
| condition |  |  |  | NA |
| Taxrate | 7.67 | 7.69 | 6.29 | NA |
| R1 A | - | - | - | 4192 |
| R1 B | - | - | - | 5219 |
| R 2 | - | - | - | 628 |
| S 2 | - | - | - | 19 |

**CONCLUSION**

According to the results and following from the hypothesis, zoning does affect residential property value. R1 A zoning, which is more restrictive than other types of zoning, tends to have the largest effect on residential property value. All significant zoning options affect the value of residential property positively raising their value except for S2. The effect of zoning on the value of residential property can be explained on the demand side by restriction of demand to a certain use, which could decrease the value. Zoning may be seen by residents as a protection of their rights while it decreases options for developers. On the supply side, zoning restricts the supply of residential property according to categories of use, which could raise value. Much consideration needs to be taken into account when public entities zone residential property; the effects could be helpful or disastrous in this research the zone of S2 has proven to be disastrous. In future studies, one needs to account more for the subjectivity of the assessed value of the properties. Each assessor gives his or her professional opinion on the value of the property, but it is only an opinion; a buyer may value the property more or less than the assessor. Moreover, since this study focuses only on property in the city of Normal, Illinois it may not be applicable to some other areas of the world that are significantly different in their economic characteristics. For example, city of Normal is economically very different from Mumbai, Chicago, Melbourne, Moscow. Moreover, better measures need to be developed to accurately separate the effects of the differentiation of provision of public services within and outside a municipality. This research was conducted to help better know the effects of zoning when combined with other physical variables which would help future researchers to develop much better model for more simplified assessment.

## AUTHOR INFORMATION

**Prathamesh Muzumdar** is a Research Associate in the Department of Marketing, Illinois State University. He has published in engineering journals and recently published a book entitled *Inflation in India - Occurrence, Causes and Measures*.

16. Rosen, S (1974), "Hedonic Prices and Implicit markets: Product Differentiation in Pure Competition" *Journal of Political Economy*, 82(1):34-55.

**NOTES**